\documentclass[onecolumn,authoryear]{els-mrw} 
 
\usepackage{amsmath,amssymb,amsfonts,amsthm,makeidx,graphicx}
\usepackage{txfonts}
\usepackage{helvet}

\usepackage{tcolorbox}

\begin{document}

\chapter{Observing Binaries}\label{chap1}

\author[1,2]{Hugues Sana}%
\author[1]{Jasmine Vrancken}%

%\author[1,2]{Third Author}%
\address[1]{\orgname{Institute of Astronomy}, \orgdiv{KU Leuven}, \orgaddress{Celestijnlaan 200D, 3001 Leuven, Belgium}}
\address[2]{\orgname{Leuven Gravity Institute}, \orgdiv{KU Leuven}, \orgaddress{Celestijnlaan 200D box 2415, 3001 Leuven, Belgium}}

\articletag{Chapter Article tagline: draft}

\newcommand{\degr}{$^\mathrm{o}$}
\newcommand{\kms}{km\,s$^{-1}$}
\newcommand{\msun}{M$_\odot$}

\maketitle

\begin{glossary}[Glossary]
\term{Apastron:} Point of furthest separation between the components of a binary system. \\
\term{Astrometric binaries:}  Binary systems that are observed due to the wobble in the proper motion direction of one of the components. \\
\term{Auxiliary circle:} Virtual circle whose diameter is given by the major axis of the orbit. \\
\term{Barycentric orbit:} Physical orbit around the center of mass of the system. \\
\term{Binary/Multiplicity fraction:} The fraction of a population of stellar objects that are in binary/multiple systems. \\
\term{Binary star (or binary system):} A dynamical system formed by two stars orbiting around their common center of gravity under their mutual gravitational attraction. \\
\term{Companion fraction:}  The average number of (stellar) companions in a population of stellar objects. \\
\term{Eclipsing binaries:} Binary systems that periodically pass in front of each other, creating observable dips in their brightness. \\
\term{Ellipsoidal variables:} Binary systems whose light curve shows signature of  tidal deformation of one or both components. They are not eclipsing binaries in the strict sense of the definition, since the light from one of the components is not necessarily blocked by the other component (but some systems display both eclipses and ellipsoidal variations). \\
\term{Periastron:} Point of closest approach between the components of a binary system. \\
\term{Primary Star:} Main reference star in the binary system (arbitrary choice, but often the most massive or the brightest). \\
\term{Radial velocity:} Component of a velocity vector projected along the observer's line of sight \\
\term{Relative orbit:} Apparent orbit of one star in the framework of the other star. \\
\term{Secondary Star:} Stellar companion to the primary star. \\
\term{Spectroscopic binaries:}  Binary systems for which their spectra display indication of binarity, either through Doppler shifted spectral lines or through multiple spectral components, or both. \\
\term{Visual doubles:} Two stars physically unrelated but coincidentally sharing a similar line of sight so that they appear at close angular separation when projected on the celestial sphere.
\end{glossary}

\begin{glossary}[Nomenclature]
%\begin{tabular}{@{}lp{34pc}@{}}
\begin{tabular}{ll|ll|ll|ll}$a$ & Semi-major axis       & G & Gravitational constant  & {\bf P} & Periastron            	& $\gamma$ & Systemic radial velocity \\       
{\bf A} & Apastron          & $i$ & Orbital inclination   & $q$ & Mass ratio                	& $\lambda$ & Wavelength \\		       
$b$ & Semi-minor axis       & $K$ & Semi-amplitude        & $\overline{r}$ & Position vector	& $\vartheta$ & True anomaly \\		       
$c$ & Speed of light        &     &of the RV curve        & $R$ & Stellar radius            	& $\mu$ & Reduced mass \\	       
CF & Companion fraction     & $M$ & Stellar mass          & $v$  & Radial velocity (RV)     	&  $\rho$ & Angular separation   \\	          
$e$ & Eccentricity          & $\mathcal{M}$ & Mean anomaly& SB &Spectroscopic Binary        	& $\phi$ & Telescope diameter  \\		          
$E$ & Eccentric anomaly     & MF & Multiplicity fraction  & $t$ & Time                      	& $\varphi$ & Orbital phase \\         
EB & Eclipsing Binary       & $N_+$ & Ascending node      & $T$ & Time of periastron passage	& $\omega$ & Argument of the periastron \\   
{\bf F} & Focus             & $N_{-}$ & Descending node   & $T_0$ & Time of primary conjunction & $\Omega$ & Longitude of  \\
$\mathcal{F}$ & Stellar flux& $P$ & Orbital period        & $T_\mathrm{eff}$ & Effective temperature &  & the ascending node
%$a$ & semi-major axis       &$M$ & Stellar mass              &$T_\mathrm{eff}$ & Effective temperature\\
%{\bf A} & Apastron          &$\mathcal{M}$ & Mean anomaly         &$\gamma$ & Systemic radial velocity \\
%$b$ & semi-minor axis       &$N_+$ & Ascending node          &$\lambda$ & Wavelength \\
%$c$ & speed of light        &$N_{-}$ & Descending node       &$\vartheta$ & True anomaly \\
%$e$ & eccentricity          &$P$ & Orbital period            &$\rho$ & Angular separation \\
%$E$ & Eccentric anomaly     &{\bf P} & Periastron            & $\phi$ & Telescope diameter  \\
%EB & Eclipsing Binary       &$q$ & Mass ratio                & $\varphi$ & Orbital phase  \\
%{\bf F} & Focus             &$\overline{r}$ & Position vector&$\omega$ & Argument of the periastron\\
%$\mathcal{F}$ & Stellar flux&$R$ & Stellar radius            & $\Omega$ & Longitude of the ascending node \\
%G & Gravitational constant  &$v$  & Radial velocity (RV)     &\\
%$i$ & Orbital inclination   &SB &Spectroscopic Binary        &\\
%$K$ & Semi-amplitude        &$t$ & Time                      &\\
%    &of the RV curve        &$T$ & Time of periastron passage&\\
\end{tabular}
\end{glossary}

\begin{abstract}[Abstract]
Binary stars are dynamical systems formed by two stars that are physically bound by the gravitational force. Binary stars are privileged laboratories, allowing one to not only measure the fundamental properties of stars but also potentially change the way stars live and die. Because of this, binary stars have continuously played a central role in astrophysics. In this chapter, we focus on the observational properties of binary stars. What are the fundamental quantities that describe binaries? How do we detect and classify them? Which parameters can we constrain for different type of objects?
\end{abstract}

\begin{tcolorbox}[colback=black!20!white]
\textbf{Objectives}
\bigskip
\begin{itemize}
 \item Understand the general concepts and key parameters that define a binary star system.
 \item Describe intrinsic and orbital parameters of binary systems and explain their significance.
 \item Derive and explain the fundamental equations governing binary star motion and their observable properties.
 \item Classify and identify binary systems based on observational techniques and their general characteristics.
 \item Illustrate the observed parameter space of binary star systems.
\end{itemize}
\end{tcolorbox}

%\section{Level 1}
%\subsection{Level 2 heading in sentance case}\label{chap1:subsec1}
%\subsubsection{Level 3 heading in sentance 
%\paragraph{Level 4 heading in sentance case}
%\subparagraph{Level 5 heading in sentance case}%
%\subsubparagraph{Level 6 heading in sentance case}%
%\subsubsubparagraph{Level 7 heading in sentance case}%

\section{Introduction}\label{s:intro}
About half the stars with a mass ($M$) similar to that of our Sun ($M \sim 1$~\msun), and most of the massive stars ($M>10$~\msun), have one or more stellar companions \citep{Duchene2013,Offner2023}. Not only are these multiple stars frequent, but they also play a primordial role in astrophysics. They provide direct constraints on fundamental properties of stars, evolve in a multitude of (binary) evolution pathways, and produce some of the most intriguing stellar physics \citep{Tauris2023,MarchantBodensteiner2023,Xuefei2024}. Understanding the formation and evolution of binary and multiple systems is therefore essential for developing a complete and accurate picture of stellar astrophysics. One of the key advantages of studying binaries is that their orbital motion reveals information about stars that cannot be obtained from single-star observations. In this chapter, we focus on the fundamental principles and methods that allow astronomers to detect and characterize binaries \citep{Hilditch2001, budding2022}. We focus on systems formed by two normal stars governed by Newtonian dynamics, with the goal to provide an undergrad-level introduction to the vastness of the field of multiple stars.

\section{Multiplicity}
\begin{figure}[b]
    \centering
    \includegraphics[width=0.6\linewidth]{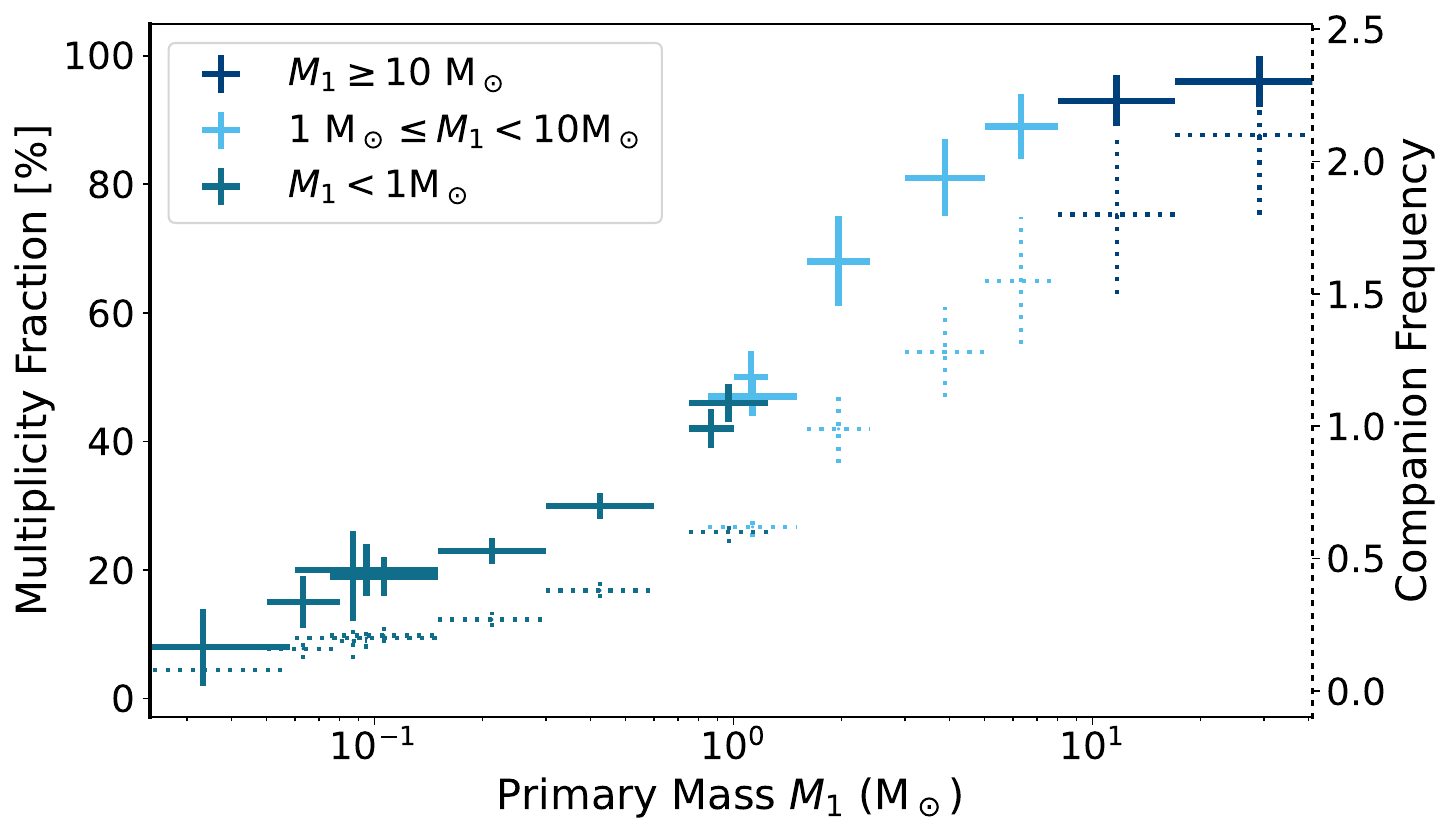}
    \caption{Multiplicity fraction and companion frequency of brown dwarfs and main-sequence stars as a function of primary mass. The references for the data points are listed in Table 1 in \citep{Offner2023}. The bold markers (left y-axis) represent the multiplicity fraction, while the dotted markers (right y-axis) correspond to the companion frequency. \textit{Figure adapted from \citet{Offner2023}; see their Table 1 for references.}}
    \label{fig:multiplicity}
\end{figure}

Recent studies on stellar multiplicity have established that most stars have one or more companions. While approximately half of solar-type stars are found in binary systems \citep{raghavan2010, tokovinin2014}, almost all massive stars \citep{Sana2012, sana2014} are part of a binary system (Fig. \ref{fig:multiplicity}). The multiplicity fraction (MF) is defined as: 
$$ MF = \frac{B + T + Q^+}{S + B + T + Q^+}$$ where $S$, $B$, $T$ and $Q^+$ represent the number of single, binary, triple and higher-order star systems, respectively. 
Another key statistics is the average companion frequency (CF), given by:
$$ CF \approx \frac{B + 2T + 3Q^+}{S + B + T + Q^+}$$
Both MF and CF exhibit an increasing trend as a function of the primary mass $M_1$ (Fig. \ref{fig:multiplicity}).

\section{General concepts}\label{s:general}

Let us consider a binary system formed by two stars of current mass $M_1$ and $M_2$, respectively. Astronomers often choose initially more massive stars as the {\bf primary star} (the star denoted '1') while its companion is called the {\bf secondary star} (denoted '2'). However, other choices can be made depending on the context or the goal of the study (e.g., the most luminous star or the currently most massive star). The choice of primary and secondary stars does not need to be unique and is mostly used as a convention to identify each of the stars of the binary system. While the equations in this text are independent of which star is which (e.g., which is the most massive or the most luminous), it is, however, important to clearly communicate the definition used to avoid confusion in the identification of the stars. Here, for example, we will assume that the primary is currently the most massive star of the pair, i.e. $M_1\geq M_2$.

\begin{figure}[t]
\centering
\includegraphics[width=.44\textwidth]{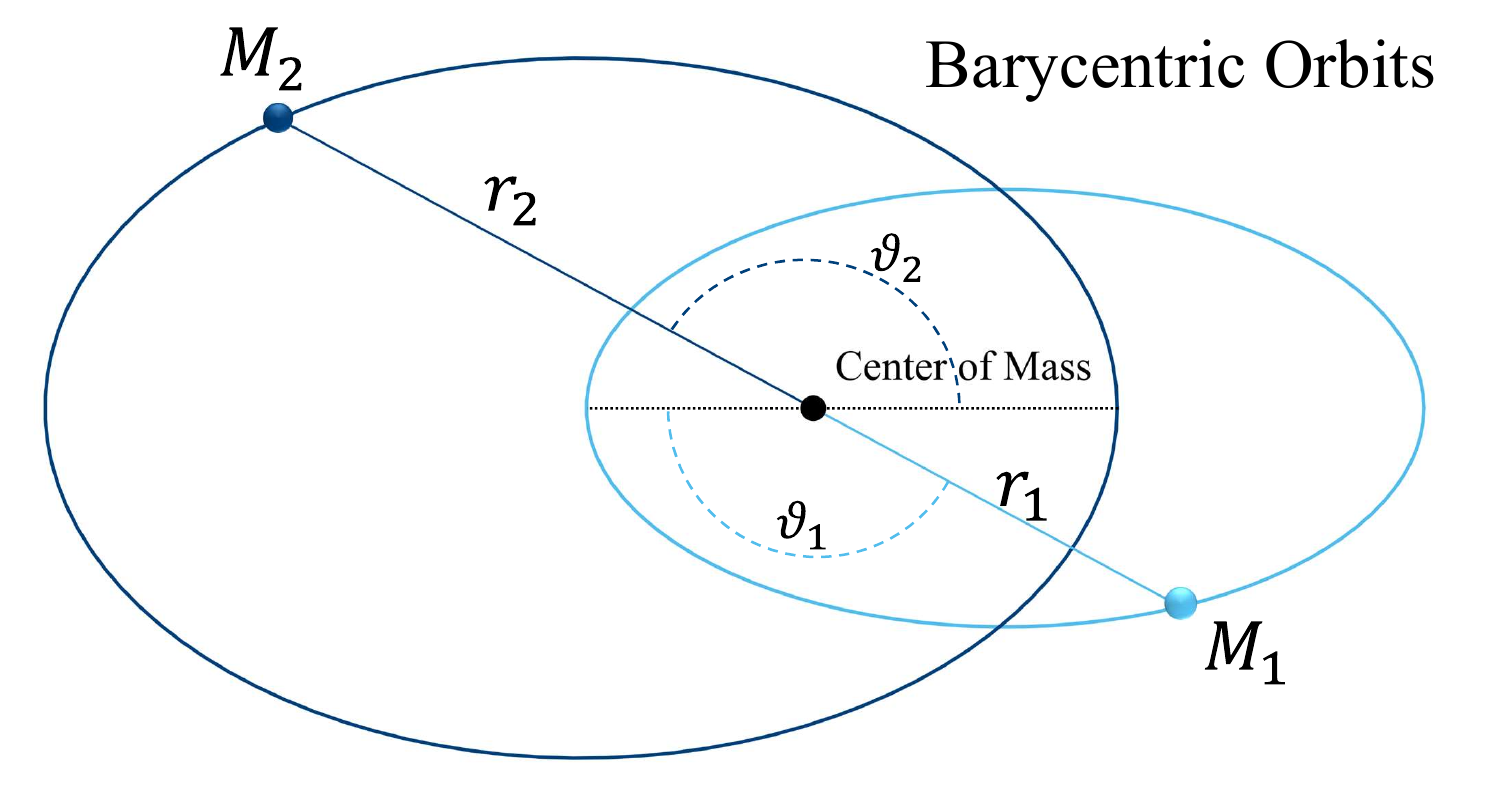}
\includegraphics[width=.55\textwidth]{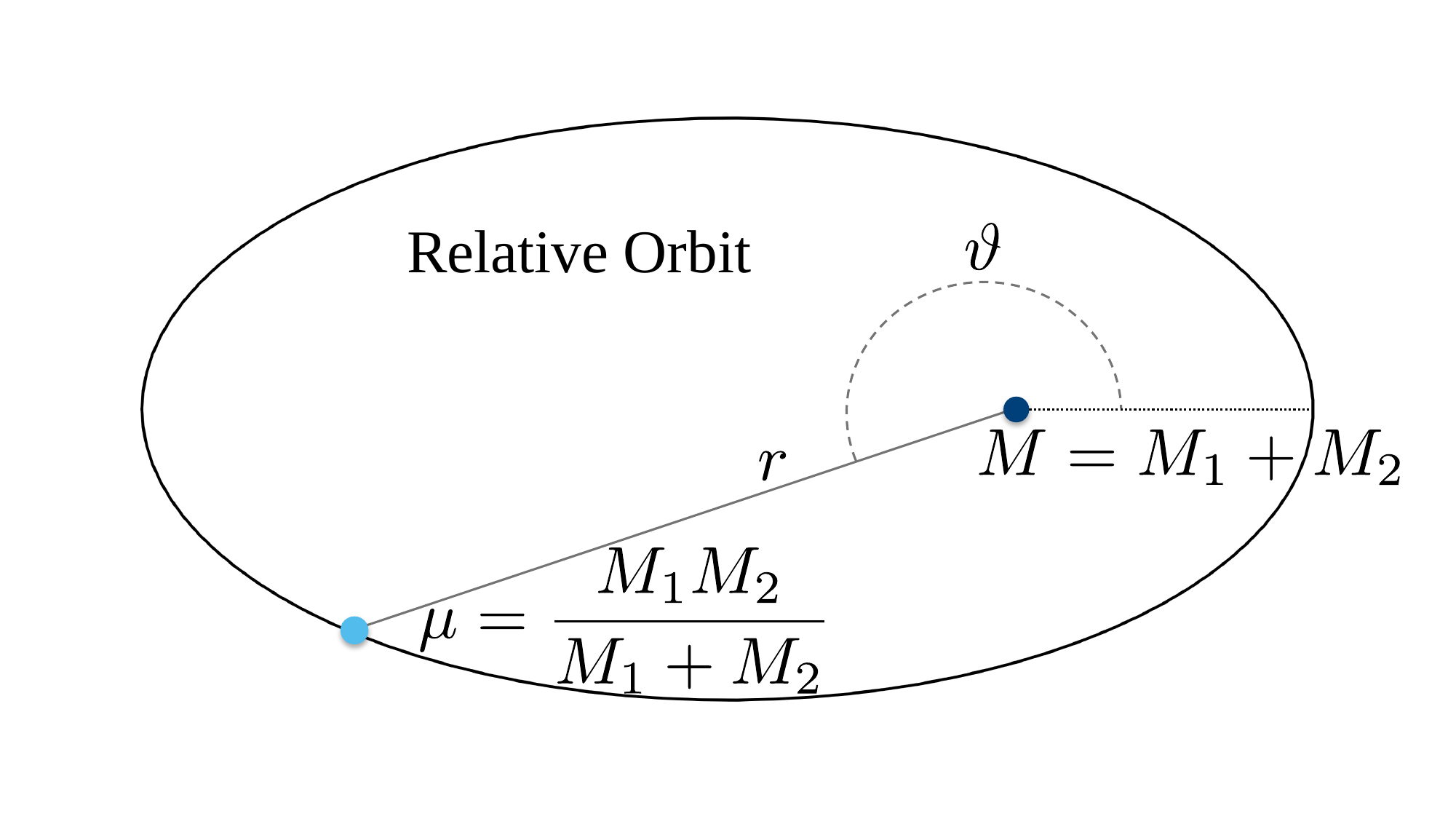}
\caption{Sketches displaying (Left) the barycentric orbits and (Right) the relative orbit of a binary system with primary and secondary masses $M_1$ and $M_2$. $\overline{r}_1$, $\overline{r}_2$ and $\overline{r}$ are the primary, secondary, and relative position vector ($r=r_1 + r_2$), $M=M_1+M_2$ is the total mass, $\mu$ is the reduced mass and $\vartheta$ are the true anomalies of the different orbits. 
}
\label{fig:bary_orbits}
\end{figure}

In a binary system, each of the two stars travels on their own {\bf elliptical orbits}. The trajectories of the primary and secondary stars are distinct but tightly related to one another (Fig.~\ref{fig:bary_orbits}): both are elliptical trajectories which occur in the same plane, have aligned major axes, and are characterized by the same {\bf eccentricity ($e$)}. They share a common \textbf{focus} ({\bf F}) which coincides with the location of the center of mass of the system. The timing of the orbital motion of both stars is also synchronized so that both stars approach and move away from the focus at the same time. The points of closest and largest separations to the focus are called the {\bf periastron} and {\bf apastron} ({\bf P} and {\bf A} on Fig.~\ref{fig:orbit}, respectively). At periastron, the separation of star $j$ to the focus is given by $a_j(1-e)$; at apastron, it is $a_j(1+e)$. At periastron, the stars are not only the closest to the focus, but the distance that separates them is also reaching a minimum, given by $a(1-e)$, with $a=a_1+a_2$. At apastron, the situation is reversed and the separation between the stars reaches a maximum, given by $a(1+e)$. The separation between the two stars varies thus by a factor of three for $e=0.5$; it varies by a factor of 19 for $e=0.9$.  

In eccentric orbits ($e\neq0$), the {\bf time of periastron passage ($T$)} is generally used as a reference time to measure the progression of the orbital motion in the time domain, while the location of the periastron is used as a reference direction to measure the angular progression of the stars on their orbit. To this purpose, we define the {\bf true anomaly ($\vartheta$)}, which is the angle formed by the direction of the periastron ({\bf $\overline{\mathrm{OP}}$}, see Fig.~\ref{fig:orbit}) and the instantaneous position of the stars ($\overline r$), measured from the focal point ({\bf F}, see Fig.~\ref{fig:orbit}). In circular orbits ($e=0$), the notion of periastron disappears as $r=a$ at all point in the orbit, and it is customary to adopt the time of primary conjunction ($T_0$) as reference time.

\section{Parameters of a binary system}\label{s:param}

In the reference frame of the center of mass of a binary system formed by two stars of masses $M_1$ and $M_2$, the shape and timing of the orbital motion are characterized by a set of six {\bf orbital parameters}. 
The {\bf orbital period ($P$)} (or equivalently, per Kepler's third law, the sum of {\bf semi-major axes $a_{1}+a_{2}$}) and {\bf eccentricity ($e$)} provide the size and shape of the orbital trajectories. The {\bf orbital inclination ($i$)} and the {\bf  longitude of the ascending node ($\Omega$)} define the position of the plane of the orbit with respect to the observer. Finally, the {\bf argument of the periastron ($\omega$)} provides the orientation of the elliptical trajectory with respect to the line of the nodes. As discussed above, the argument of the periastron ($\omega$) further provides a reference direction to measure the true anomaly ($\vartheta$) while the time of periastron passage ($T$) give the moment at which the system passes through the reference configuration (i.e., $\vartheta(t=T)=0$), with $t$, the time.  Equivalent times of periastron passage differ by an integer number of times the orbital period ($T + n P$, with $n \in \mathbb{N}$).

\begin{figure}[t]
\centering
\includegraphics[width=.4\textwidth]{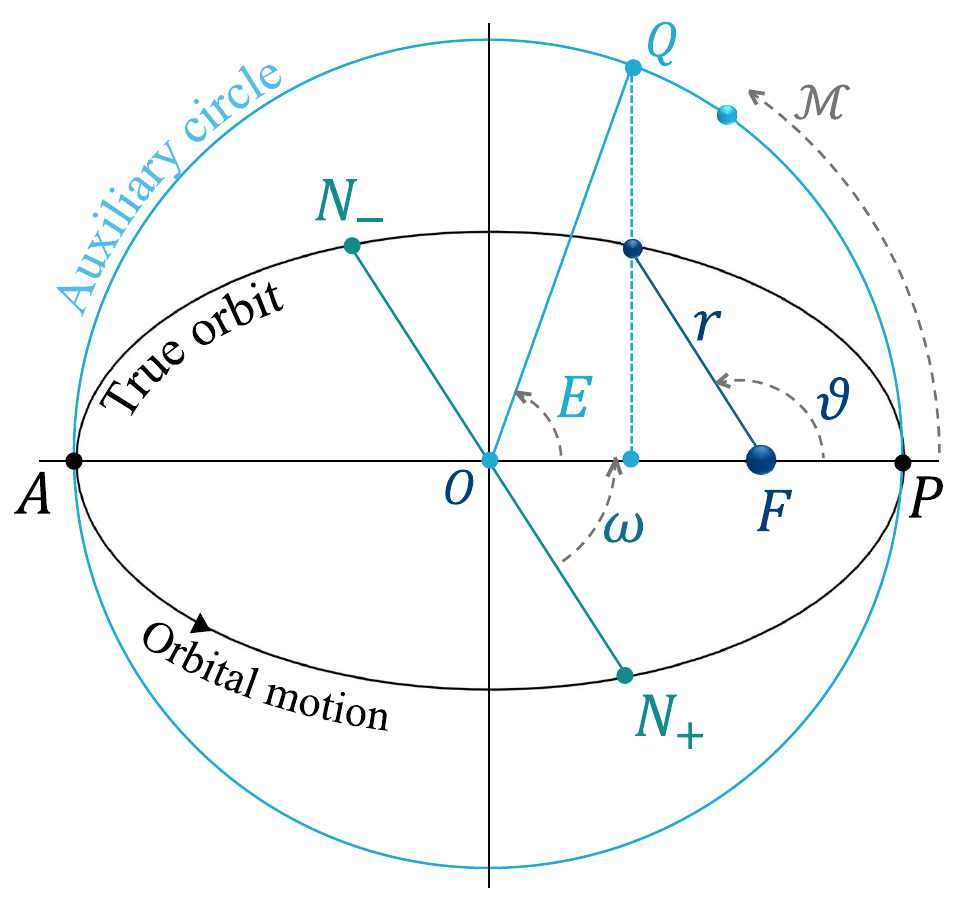}
\includegraphics[width=.55\textwidth]{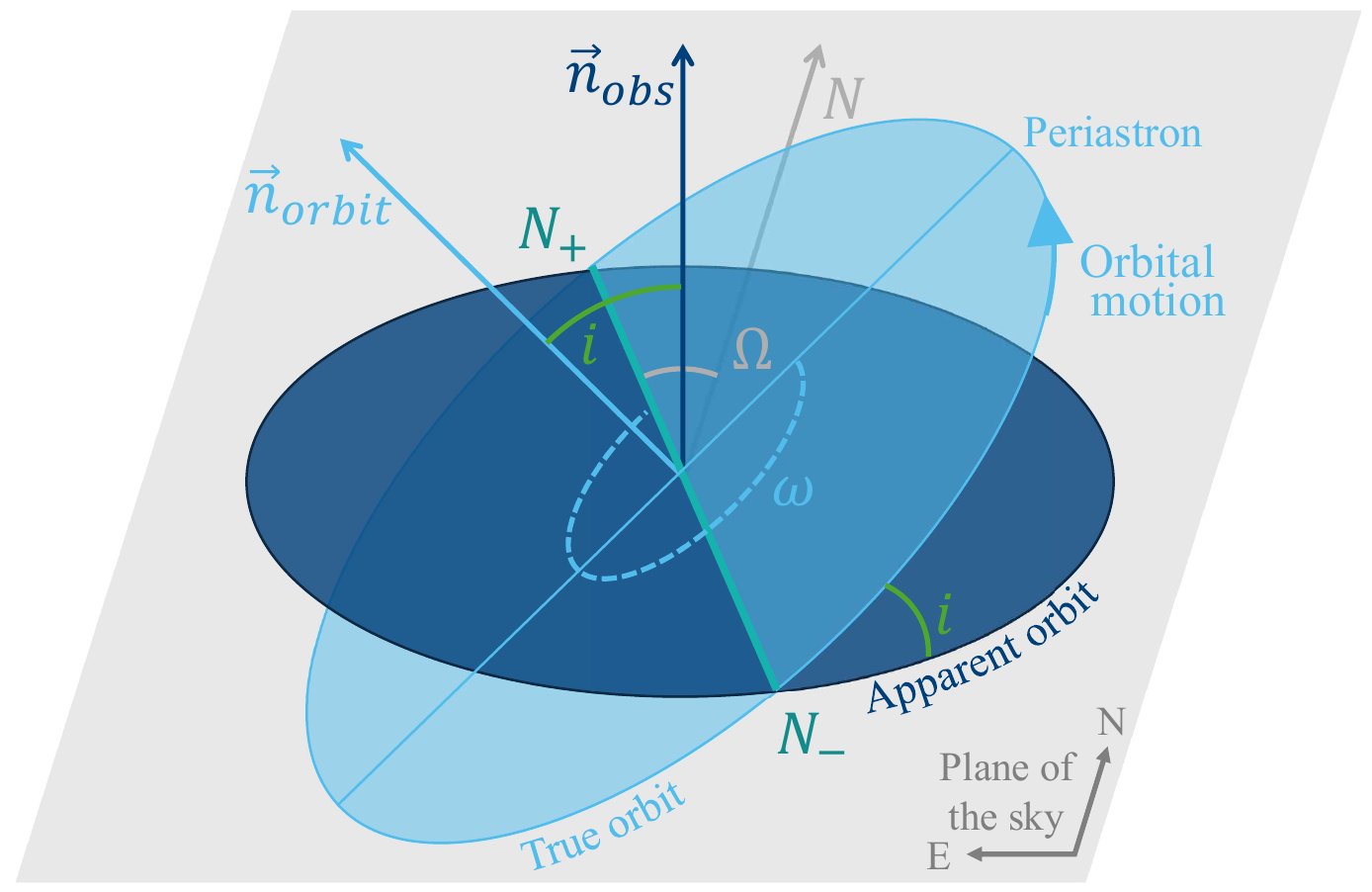}
\caption{(Left:) The orbital trajectory  and auxiliary circle, and the three anomaly angles ($\vartheta$: true anomaly, $E$ eccentric anomaly and $\mathcal{M}$ mean anomaly) used to describe the orbital motion along the orbit.  (Right:) Three-dimensional geometry of an orbit and the location of the Kepler's angles used to define it ($i$: orbital inclination, $\Omega$: longitude of the ascending node, and $\omega$: argument of periastron. }
\label{fig:orbit}
\end{figure}

The three angles $i$, $\Omega$, and $\omega$ fully describe the  orientation of the orbit in the three-dimensional space and are sometimes referred to as {\it Kepler's angles}. They play a crucial role in defining the observational properties of binary systems. Given their importance, some further concepts are needed (see sketch in Fig.~\ref{fig:orbit}, right):
\begin{enumerate}
    \item[-] the {\bf plane of the orbit} is the plane in which the orbital motion of the binary system takes place;
    \item[-] the {\bf plane of the sky} is the tangential plane to the celestial sphere at the position of interest. It is always at 90$^\circ$ compared to the line of sight;
    \item[-]  the {\bf line of nodes} is formed by the intersection of the plane of the orbit with the plane of the sky. The intersection of the orbital trajectory with the line of nodes defines the ascending and descending nodes of the orbit. The {\bf ascending node ($N_+$)} is the node at which the primary star is moving away from the observer while passing through the plane of the sky, while the primary stars will approach  the observer while passing through the {\bf descending node ($N_-$)}.
    \end{enumerate}
    Using these concepts, we can refine our definitions of the Kepler's angles:
    \begin{enumerate}

   \item[-] the {\bf orbital inclination ($i$)} is the angle between the plane of the sky and the plane of the orbit. It is also the angle between the line of sight of the observer and the norm to the plane of the orbit.  The inclination will have values between $-\pi/2$ and $+\pi/2$;
    \item[-] the {\bf longitude of the ascending node ($\Omega$)} is the angle  between the true North and the ascending node. It is measured in the plane of the sky from North to East and takes values between 0 and $2\pi$;
    \item[-] the {\bf argument of the periastron ($\omega$)} is the angle between the ascending node ($\overline{\mathrm{ON_+}}$) and the direction of periastron ($\overline{\mathrm{OP}}$). It is measured in the plane of the orbit, from the ascending node, and increases in the direction of the orbital motion. The argument of periastron takes values between $0$ and $2\pi$.
\end{enumerate}

Aside from these six orbital parameters, six additional quantities are required to describe the position and velocity of the center of mass of the systems in the 3D space. The three spatial coordinates are usually given by the \textbf{distance $d$} to the system, and by two angles providing the position on the celestial sphere (e.g., right ascension and declination) at a given time (e.g., at equinox J2000.0). The system velocity vector is  usually decomposed into its line of sight component -- the {\bf systemic radial velocity ($\gamma$)} --  and its tangential component or {\bf proper motion} (i.e., the velocity component in the plane of the sky). 

In the strict assumptions of a two-body problem with point-like masses, these six orbital parameters ($P, e, T, i, \omega, \Omega$) do not change over time and are fully described by Kepler's laws (see Sect.~\ref{s:eqn}). In reality, however, the stars are not isolated point sources and additional physics may produce variation over time. Examples include (i) the precession of the argument of the periastron, or apsidal motion ($\dot{\omega}=\frac{\mathrm{d}\omega}{\mathrm{d}t}$), resulting from shortcomings of the point-like assumption, especially if stars are deformed, or from relativistic corrections; (ii) a variation of the orbital period ($\dot{P}$), which may result from a change in the masses of the stars, e.g., due to stellar winds or mass transfer; and (iii) a change of eccentricity ($\dot{e}$), e.g. through tidal interaction. The presence of additional companions, e.g.\ in a triple or higher-order system, may also affect  the orbital parameters through angular momentum exchange or orbital resonance. While important, these processes are beyond the goal of the present overview.

 \section{Absolute vs. relative orbit}\label{s:orbit}
An important distinction, and a source of confusion at first among  students, is the difference between the barycentric or absolute orbits, and the relative orbits (see Fig.~\ref{fig:bary_orbits}). The importance of this difference will become more apparent in the next section, as certain observing techniques are more suited to characterize the barycentric orbits while others more easily measure relative positions or relative motions.

The  {\bf barycentric orbits} describe the motion of each of the binary component around the center of mass of the system, and is usually described in the rest frame of the center of mass. The barycentric orbits are described in polar coordinate by:
\begin{eqnarray}
    r_1(t)&=&\frac{a_1(1-e^2)}{1+e \cos \vartheta_1(t)},\\
    r_2(t)&=&\frac{a_2(1-e^2)}{1+e \cos \vartheta_2(t)}.
\end{eqnarray}
As discussed in Sect.~\ref{s:general}, each of the components of the binary system evolve on its own elliptical trajectory. Both ellipses have the center of mass in one of their focus and move synchronously such that the numerical values of $\vartheta_1(t)$ and $\vartheta_2(t)$ are identical at all times. However, we emphasize that $\vartheta_1(t)$ and $\vartheta_2(t)$ are physically different angles and are measured from their respective periastron direction (see Fig. \ref{fig:bary_orbits}, left). The latter are given by $\omega_1$
and $\omega_2$ and satisfy $\omega_2=\omega_1+\pi$.

The {\bf relative orbit} describes the change of position of one of the component (usually the secondary, faintest and/or least massive) with respect to the other component. It thus describes the evolution of the end point of the relative position vector $\mathbf{r}$ over time. Adopting a reference frame centered on star $1$, the relative ellipse is then given by:
\begin{eqnarray}
\label{eq:r}
        r(t)&=&\frac{a(1-e^2)}{1+e \cos \vartheta(t)},
\end{eqnarray}
where $a=a_1+a_2$ and $\vartheta(t)=\vartheta_1(t)=\vartheta_2(t)$ at all times (though again they are physically different angles). The relative and barycentric orbits  share the same period, eccentricity, time of periastron passage, and orbital inclination. Importantly, in the present convention, the argument of the periastron of the relative orbit is that of the secondary star $\omega=\omega_{2}=\omega_{1}+\pi$ given the relative orbit describes the location of the secondary with respect to that of the primary. Graphically, while the relative orbit describes the position of the secondary with respect to the primary, it actually corresponds to the Kepler's equations for a fictitious binary system where the primary star has a mass $M$ equal to the total mass $M_1+M_2$ of the true system and the secondary star as a reduced mass of $\mu=\frac{M_1 M_2}{M_1+M_2}$ (see Fig.~\ref{fig:bary_orbits}, right).

\section{Fundamental equations}\label{s:eqn}
The orbital motion of a binary problem is no more than a 2-body problem. It is a generalized version of the laws that  Kepler derived for the solar system where we consider this time two objects with non-negligible masses. These equations can be easily derived from first principles in classical mechanics. The elliptical shape of the orbit is given by:
\begin{equation}
\frac{x^2}{a^2}+\frac{y^2}{b^2}=1,
\label{eq:cartesian}
\end{equation}
where $(x,y)$ are the Cartesian coordinates in the plane of the orbit, and $a$ and $b$ are the semi-major and semi-minor axis of the orbit. The eccentricity is then given by $e=\sqrt{a^2-b^2}$.  Traditionally, astronomers rather work in polar coordinates, $(r,\vartheta)$, with the center of coordinates at the focus {\bf F} of the ellipse. Under these assumptions, the ellipse equation becomes:
\begin{equation}
    r(t)=\frac{a(1-e^2)}{1+e \cos \vartheta(t)} \label{eq:K1}
\end{equation}

where $\mathbf{r}(t)$ is the position vector ($r=|\mathbf{r}|$, its norm), and $\vartheta(t)$ the true anomaly. Eqs.~\ref{eq:cartesian} and  \ref{eq:K1} are the mathematical form of Kepler's first law.

Projected on a 3D Cartesian set of coordinates in the observer's rest frame (with the $x$-axis oriented towards the celestial North and the $z$-axis parallel to the observer's line of sight), the orbital motion described by Eq.~\eqref{eq:K1} becomes:
\begin{eqnarray}
    x(t)&=&r \left( \cos \Omega \cos(\omega+\vartheta(t)) - \sin \Omega \sin(\omega+\vartheta(t)) \cos i \right) \label{eq:K1x},\\
    y(t)&=&r \left( \sin \Omega \cos(\omega+\vartheta(t)) + \cos \Omega \sin(\omega+\vartheta(t)) \cos i \right)\label{eq:K1y},\\
    z(t)&=&r \left( \sin(\omega+\vartheta(t)) \sin i \right). \label{eq:K1z}
\end{eqnarray}

The interest of using polar coordinates is that all the time dependence of the orbital motion is placed into one variable, $\vartheta$. The other parameters of the right-hand side of Eq.~\eqref{eq:K1} ($a$ and $e$) are indeed constant for a given orbit. 

The true anomaly $\vartheta$ is measured in the plane of the orbit from the argument of periastron in the direction of the orbital motion. To describe the dependence in time of the true anomaly, we first introduce the {\bf eccentric anomaly (E)}. Measured from the center of the orbit, $E$ is the angle POQ between the direction of the periastron ($\overline{OP}$) and that of a point {\bf Q}, which is the projection, along the semi-minor axis, of the position vector on the {\bf auxiliary circle} (see Fig.~\ref{fig:orbit}, left). The relation between $E$ and $\vartheta$ is given by: 

\begin{equation}
\tan \left(\frac{\vartheta}{2}\right)    = \sqrt{\frac{1 + e}{1 - e}} \tan \left( \frac{E}{2} \right) 
\end{equation}
while the dependence of $E$ with the time $t$  
is given by the Kepler's equation:
\begin{equation}
    E(t) - e \sin E(t) = 2 \pi \frac{(t-T)}{P} = 2 \pi \varphi(t) = \mathcal{M}(t), \label{eq:K2}
\end{equation}
where  $\varphi=\frac{(t-T)}{P}$ is the {\bf orbital phase} and $\mathcal{M}$, the {\bf mean anomaly}. The orbital phase  expresses the fraction of the orbital period that has elapsed since the reference point. The mean anomaly is the equivalent of the phase, but in units of angle. $\mathcal{M}(t)$ and $\varphi(t)$  always have a linear dependence with time, which makes them intuitive variables for planning and analyzing observations. However, in eccentric systems, neither $\vartheta$ nor $E$ vary linearly with time: they vary faster close to periastron  $\left( \vartheta(t=T)=E=0 \right)$ and slower near apastron  ($\vartheta(t=T+\frac{P}{2})=E=\pi$). This is the generalization of Kepler's area law (often referred to as Kepler's second law) to binary systems. Kepler's equation for eccentric systems does not have an analytical solution and was historically solved graphically. Nowadays, it is solved numerically. 

Finally, Kepler's law of periods (often referred to as Kepler’s third law) established a power-law relation between the orbital period and the size of the planetary orbits in the solar systems. Generalized to a two-body system with non-negligible masses, this law becomes:
\begin{equation}
    \left( \frac{P}{2\pi} \right)^2 = \frac{a^3}{\mathrm{G} M}, \label{eq:K3}
\end{equation}
where $\mathrm{G}$ is the gravitational constant, $a$ the semi-major axis of the relative orbit ($a=a_1+a_2$, see Section \ref{s:orbit}) and $M=M_1+M_2$ is the total mass of the system. Expressed in units of years, astronomical units and solar masses for $P, a$ and $M$, respectively, Eq.~\eqref{eq:K3}  becomes:
\begin{equation}
P^2 \approx \frac{a^3}{M}.\label{eq:K3b}
\end{equation}
 The relative precision of the  approximation of Eq.~\eqref{eq:K3b} to Eq.~\eqref{eq:K3} is given by the ratio of the mass of the Earth to that of the Sun, so about $3\times10^{-6}$, which is thus a very good approximation.

\begin{figure}[t]
\centering
\includegraphics[width=.75\textwidth]{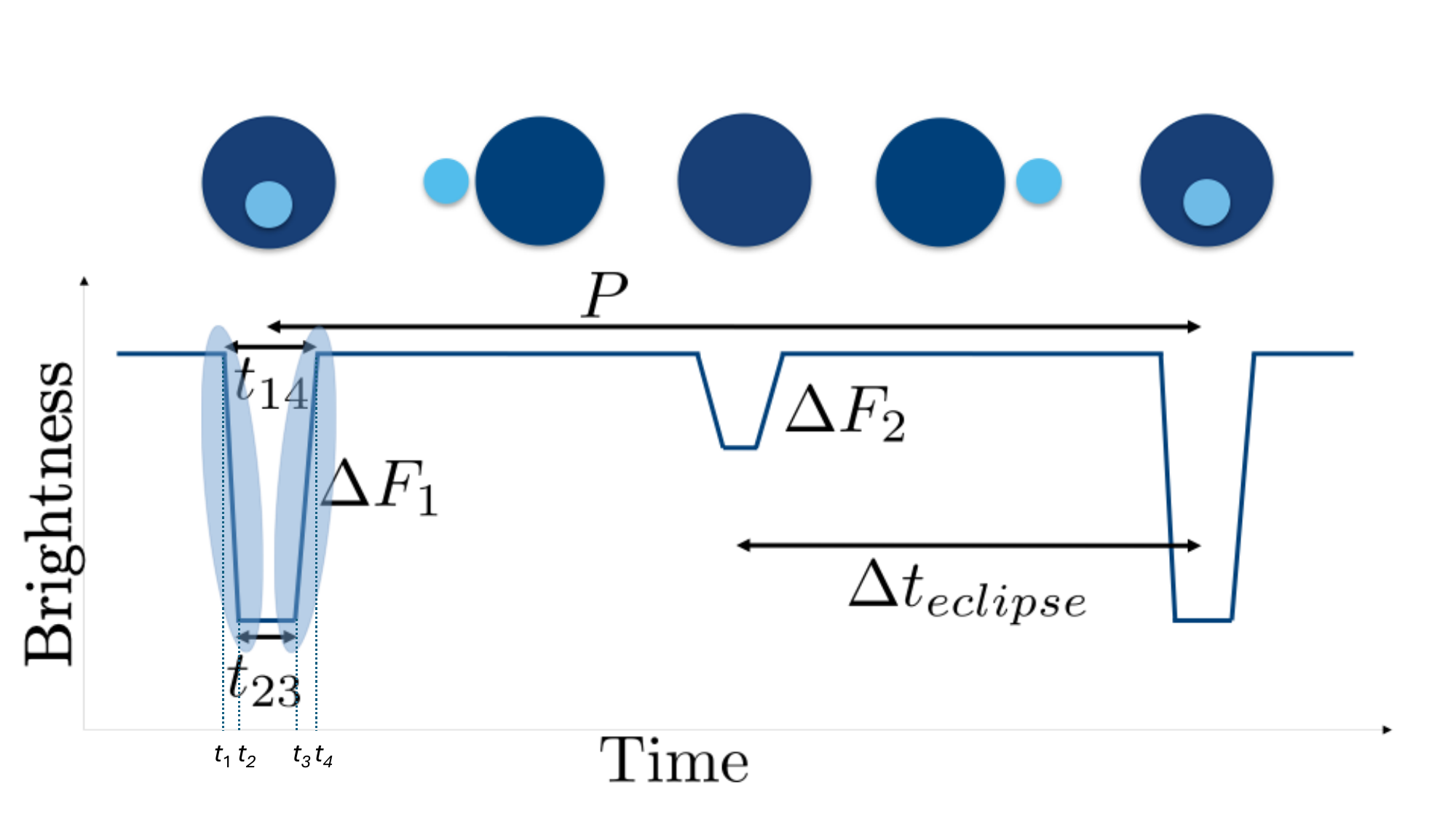}
\caption{Illustration of a transit light curve highlighting some features that reveal information about the intrinsic, orbital and geometrical parameters of the system.}
\label{fig:transit}
\end{figure}
 
% \section{Observational techniques }\label{s:det}

 \section{Observational techniques}\label{s:obs}
Observing methods can broadly be organized into two main categories. On the one hand,  {\bf statistical methods}  rely on statistical evidence to detect (candidate) binary systems. These methods often  look at properties of large samples of stars such as their brightness distribution (e.g., the distribution of stars above the main sequence in the Hertzsprung-Russell diagram), their (radial) velocity dispersion (as the orbital motion of binaries add to the velocity dispersion compared to a sample of single stars), their spatial distribution (e.g., by comparing the observed alignments of stars on the line of sight compared to statistically fortuitous alignments), or their joint proper motion.

On the other hand, {\bf direct detection methods} attempt to detect the signature of the orbital motion to prove that an object cannot be a single star. To do so, these methods will search for evidence of the orbital motion in the properties of at least one of the components. More often than none, they will gather time-resolved observations and  interpret their temporal variations in the context of a binary motion. Ideally, the data would also be compared to single star models, revealing a (much) poor(er) fit to the data than when using a binary model.  %For example, the Gaia satellite used the variation of the position of the photocenter of a binary system to detect the orbital motion while interferometry looks at the change of 

Both direct and statistical approaches can rely on a variety of observational techniques, including photometry, spectroscopy and imaging. Usually, they probe for binary signatures in the object flux, position, velocity and/or acceleration (e.g., change of acceleration towards/away from periastron, or black hole and neutron star mergers that are detected through their gravitational wave emission resulting from the acceleration imposed on the masses by gravity).
In the following, we will focus on direct detection methods only. Categorizing these methods even further by using their main scrutinized quantity, one can distinguish:
\begin{enumerate}
    \item[-]{\bf Photometric methods}, which search for variations of the light flux;
    \item[-] {\bf Astrometric methods}, which search for variations of the absolute or relative positions;
    \item[-]{\bf Spectroscopic methods}, which search for variations in the spectral domain, often through Doppler effect, and hence tend to focus on line-of-sight velocity information;
    \item[-]{\bf Imaging methods}, which search for direct detection of a companion's light flux, with or without the possibility to also monitor changes in its position.
\end{enumerate}
Astronomers like to categorize objects, and the field of binaries is no exception. Several classification schemes exist, e.g. by using the  physical radii to separation ratio (see Chapter \textit{Evolution of binary stars}), or by comparing the binaries under consideration with a prototype object that shares similar properties. As this article focuses on the observational aspects, we will  adopt a widely used classification scheme, which focuses on the detection signature. In the following we briefly review some of the most popular types of binaries, emphasizing their specificity, the orbital properties that can be constrained from their study and areas of blind spots.

\subsection{Eclipsing binaries}
 Eclipsing binaries (EBs) are binary systems that are detected through the presence of {\it eclipses} (also called {\it transits}). These occur when one of the stars passes in front of the other in our line of sight, such that it blocks (part of) the light of its companion (Fig.~\ref{fig:transit}). This periodic decrease in the flux received from the system is called an eclipse or transit. Depending on the interplay between orbital separation, inclination and the radii of the stars, and, in non-circular system, the eccentricity and periastron argument, one may observe one or two eclipses (single or double-eclipse systems), or not at all (non-eclipsing binaries). As an underlying principle, the inclination needs to be favorable so that the projections of the stellar disks on the plane of the sky will overlap at a certain time during the orbit. Limiting ourselves to circular systems, this can be expressed as a set of conditions linking the inclination $i$ and the stellar radii $R_1$ and $R_2$ relative to the orbital separation $a$: 
\begin{align*}
    |\cos i| \leq \frac{R_1 - R_2}{a} \qquad &\text{Full eclipse,} \\
    \frac{R_1 - R_2}{a} \leq |\cos i| \leq \frac{R_1 + R_2}{a} \qquad &\text{Grazing eclipse,} \\
    |\cos i| \geq \frac{R_1 + R_2}{a} \qquad &\text{No eclipse.}
\end{align*}

These equations indicate that large (relative) radii, tight separations and high inclination values are favorable configurations, while low-inclination and/or wide binaries will not produce eclipses. Similarly, the  probability of eclipses decreases with increasing orbital separations, hence orbital periods (Eq.~\eqref{eq:K3}), so that the vast majority of known eclipsing binaries have orbital periods of a few weeks at most. Given enough flux measurements to sample the light curve of the system, the shape and depth of transits in eclipsing binaries provide important information about the nature of the stars in the system. Some useful characteristics of the light curves are illustrated in Fig. \ref{fig:transit}:

\begin{figure}[t]
\centering
\includegraphics[width=.6\textwidth]{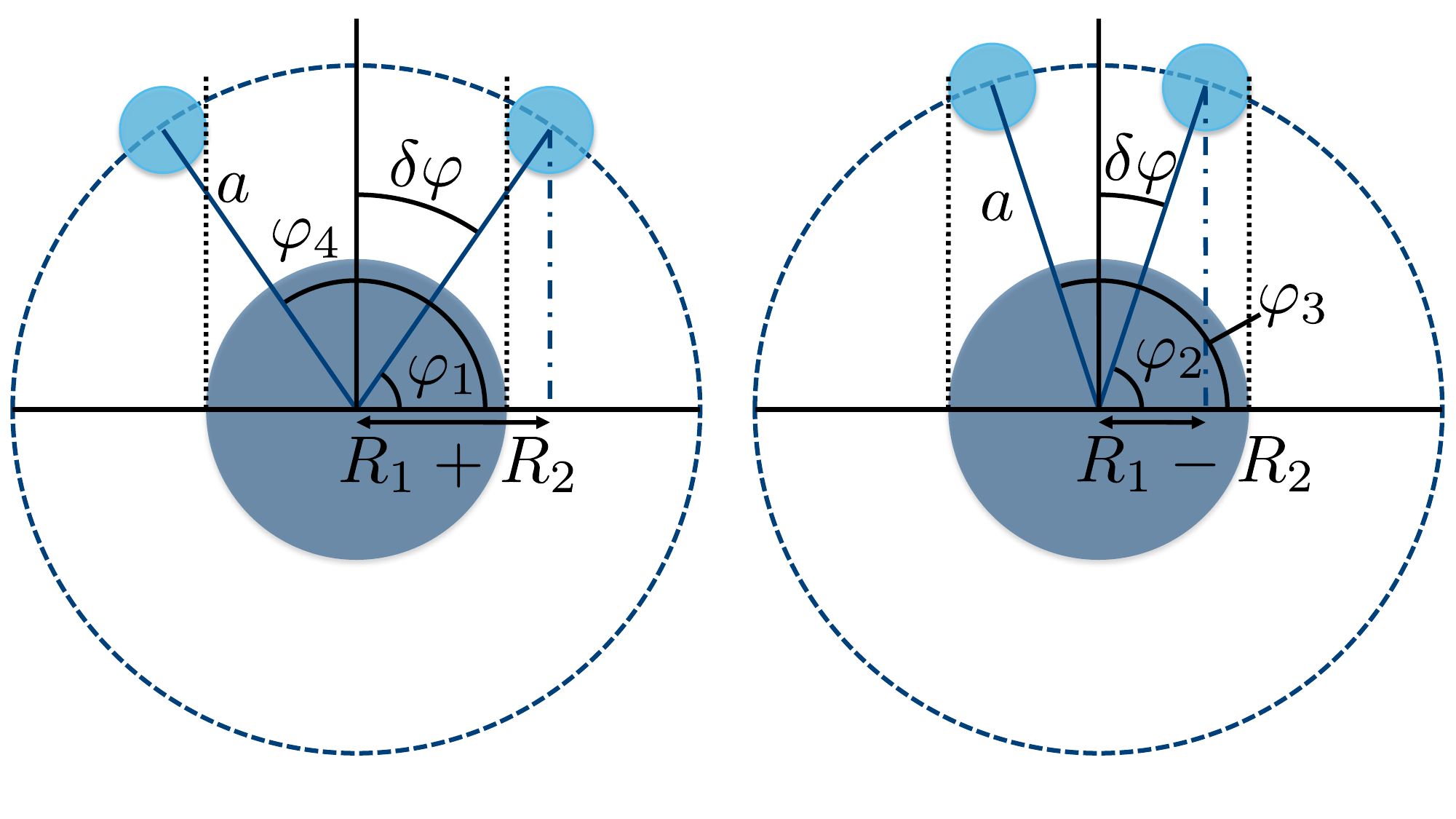}
\caption{Illustration of the geometry and angles used to relate the stellar radii to the transit duration, using the angles of ingress and egress of the eclipse.}
\label{fig:duration}
\end{figure}

\begin{enumerate}
    \item[-] The separation between the primary dips in the light curve provides a measure of the orbital period $P$ of the binary;
    \item[-]  The separation between the primary and secondary transits of double-eclipsing (EB2) systems is $P/2$ for circular orbits. Deviations from such timing is a direct evidence of eccentricity. However, depending on the orientation of the orbit with respect to the observer, not all eccentric EB2 systems can be revealed that way;
    \item[-] The total duration of the eclipse $t_{14}$ and the duration of totality $t_{23}$, provide a measure of  $R_1/a$ and $R_2/a$, the  relative size of the stars  to the (instantaneous) separation. For circular orbits, the geometry shown in Fig. \ref{fig:duration} results in:
\begin{eqnarray}
\frac{\pi}{P} t_{14} &=& \frac{\sqrt{(R_1 + R_2)^2 - (a \cos i)^2}}{a},\\
\frac{\pi}{P} t_{23} &=& \frac{\sqrt{(R_1 - R_2)^2 - (a \cos i)^2}}{a}.
\end{eqnarray}
\item[-] Assuming totality, the relative depths of the eclipses $\Delta \mathcal{F}_1/\mathcal{F}_\mathrm{tot}$ and $\Delta \mathcal{F}_2/\mathcal{F}_\mathrm{tot}$ provide insight into the ratio of the surface brightness and temperatures following
%The primary eclipse depth gives the flux ratio of monochromatic luminosities, while $\Delta F_2$ gives the ratio of the radii and the ratio $\Delta \mathcal{F}_1/ \Delta \mathcal{F}_2$ are correlated with the ratio of surface brightness. 
%Specifically, the depths of the primary eclipse $\Delta \mathcal{F}_1$ and the secondary eclipse $\Delta \mathcal{F}_2$ are related to the ratio of the radii and the ratio of the temperatures of the stars. The relation is given by:
\begin{eqnarray}
    \frac{\Delta \mathcal{F}_1}{\mathcal{F}_\mathrm{tot}} &=& \left( \left( \frac{R_1}{R_2} \right) ^2 + \left( \frac{T_\mathrm{eff,2}}{T_\mathrm{eff,1}} \right)^4 \right)^{-1}, \\
 \frac{\Delta \mathcal{F}_2}{\mathcal{F}_\mathrm{tot}} &=& \left( \left( \frac{R_1}{R_2} \right) ^2 \left( \frac{T_\mathrm{eff,1}}{T_\mathrm{eff,2}} \right) ^4 + 1 \right)^{-1}.    
\end{eqnarray}

\item[-] The shape of the ingress ($t_1$ to $t_2$) and egress ($t_3$ to $t_4$) contains information about the limb darkening, while the shape during the eclipse ($t_2$ to $t_3$) is highly dependent on the inclination $i$ of the system as well as $R_1/a$ and $R_2/a$.
\end{enumerate}
\vspace*{0.5cm}

While the above equations will allow to properly model the light curve of a well separated system, additional physics might complicate the signal. The shape of the light curve outside the eclipse(s) may not need to be constant and might be affected by, e.g., reflection, mutual heating and deformation of the stellar surfaces. In practice, state-of-the-art light curve modeling include meshing the 3D surface of the stars, and evaluate the flux that each cell emit in the direction of the observer. These so-called Wilson-Devinney methods \citep{WD1971} are not trivial to implement, but a variety of publicly available, professional softwares are available \citep[e.g.,][]{Prsa2018}. 

%With these equations and Kepler's third law, light curves can be modeled with given intrinsic and orbital parameters. By fitting the data points to these light curves, the parameters of the system can be derived. An example of a light curve fitting is shown in Fig. \ref{fig:transitfit} \jasmine{Is there a nice plot from someone of our team we can put here?}where additional information from the radial velocity curves is used as well.

%\begin{figure}[b]
%\centering
%\includegraphics[width=.6\textwidth]{Figures/blank}
%\caption{}
%\label{fig:transitfit}
%\end{figure}

\subsection{Spectroscopic binaries}

In spectroscopic binaries (SBs), the orbital motion produces a periodic variation on the position of the spectral lines. Such a displacement of the spectral line with respect to the laboratory (rest) wavelength provides a direct measurement of the line of sight velocity component of celestial objects.  
%The orbital motion of a binary system causes variations in radial velocity along our line of sight, depending on the system’s inclination.This radial velocity can be measured from shifts in the stars' spectral lines because of the Doppler effect. In short, the Doppler effect tells that objects moving away from the observer appear redshifted, while those moving towards the observer appear blueshifted.
The \textbf{wavelength shift $\Delta \lambda$} and the {\bf radial velocity} $v_\mathrm{r}$ are related to each other by the Doppler formula:
\begin{equation}
    \frac{\Delta \lambda}{\lambda_0} = \frac{\lambda - \lambda_0}{\lambda_0} = \frac{v_\mathrm{r}}{c}, \label{eq:doppler}
\end{equation}
where $\lambda_0$ is the {\it rest wavelength} (or {\it laboratory wavelength}) of the spectral line, $\lambda$ the observed wavelength of the centroid of the shifted spectral line, and $c$ is the speed of light.
Deriving Eq.~\eqref{eq:K1z} and using
%the radial velocity of the system can be expressed like:
%\begin{eqnarray}
%    v_\mathrm{r,1}(t) &=& \sin i [\sin (\vartheta_1(t) + \omega_1) \dot{r_1} + r_1 \cos (\vartheta_1(t) + \omega_1) \dot{\omega_1} ].
%    v_\mathrm{r,2} &=& \sin i [\sin (\vartheta + \omega_1) \dot{r} + r \cos (\vartheta + \omega) \dot{\omega} ].
%\end{eqnarray}
the expression for $r$ from Eq.~\eqref{eq:K1}, and Kepler's second law, the radial velocities of the two stars in  the system can be expressed  as:
\begin{eqnarray}
    v_\mathrm{r,1}(t) &=& \gamma + \frac{2 \pi}{P}\frac{a_1 \sin i}{ \sqrt{1-e^2}} [\cos (\vartheta_1(t) + \omega_1) + e \cos \omega_1], \label{eq:vr1_orig} \\
    v_\mathrm{r,2}(t) &=&  \gamma +\frac{2 \pi}{P} \frac{a_2 \sin i}{\sqrt{1-e^2}} [\cos (\vartheta_2(t) + \omega_2) + e \cos \omega_2], \label{eq:vr2_orig} 
\end{eqnarray}
where the factors 
\begin{equation}
K_{1,2}=\frac{2 \pi}{P} \frac{a_{1,2}\sin i}{\sqrt{1-e^2}}
\end{equation}
are  the semi-amplitudes of the radial-velocity curves and $\gamma$ is the systematic radial velocity.

\begin{figure}[t]
\centering
\includegraphics[width=.47\textwidth]{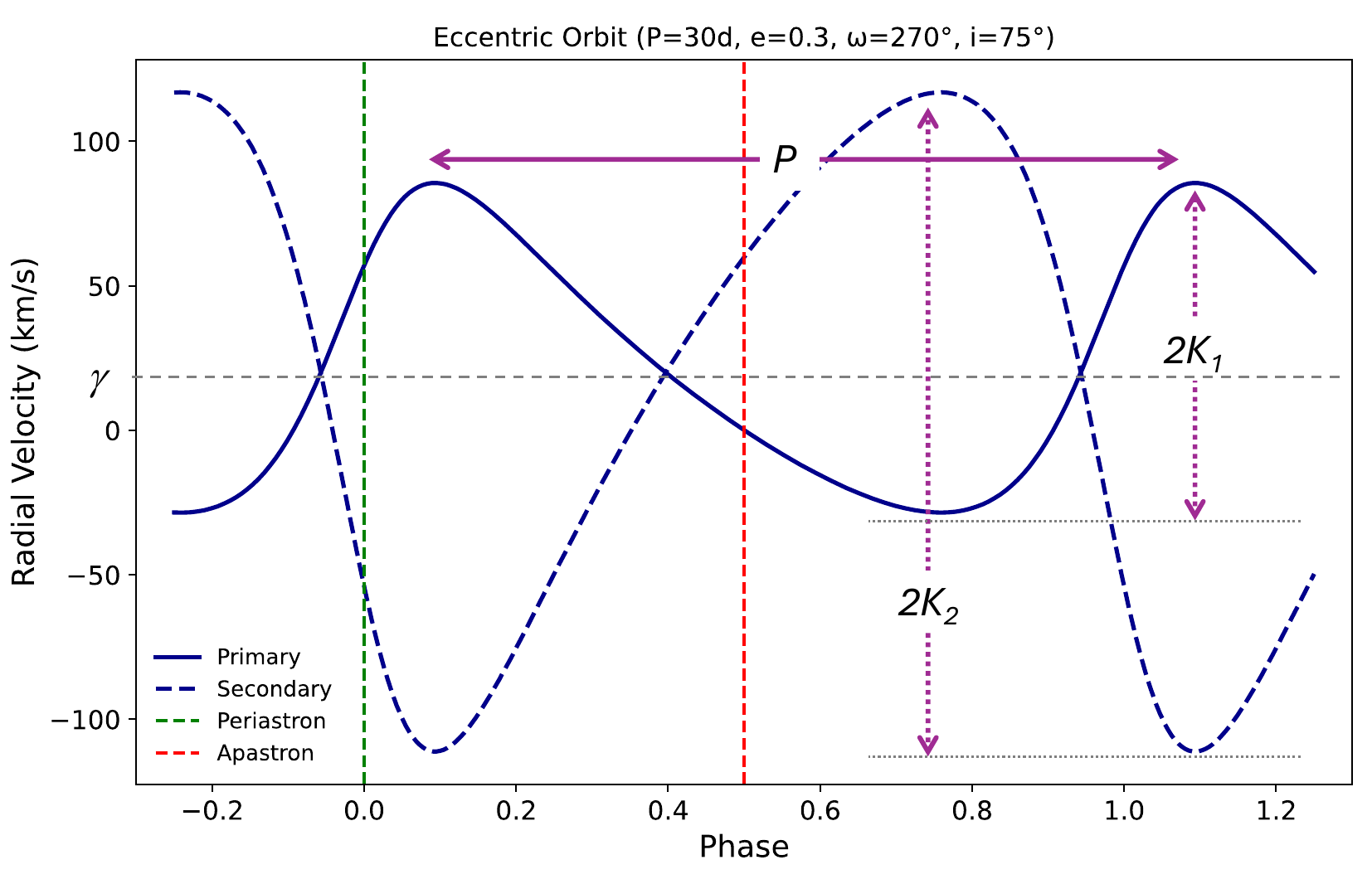}
\hspace*{0.9cm}\includegraphics[width=.47\textwidth]{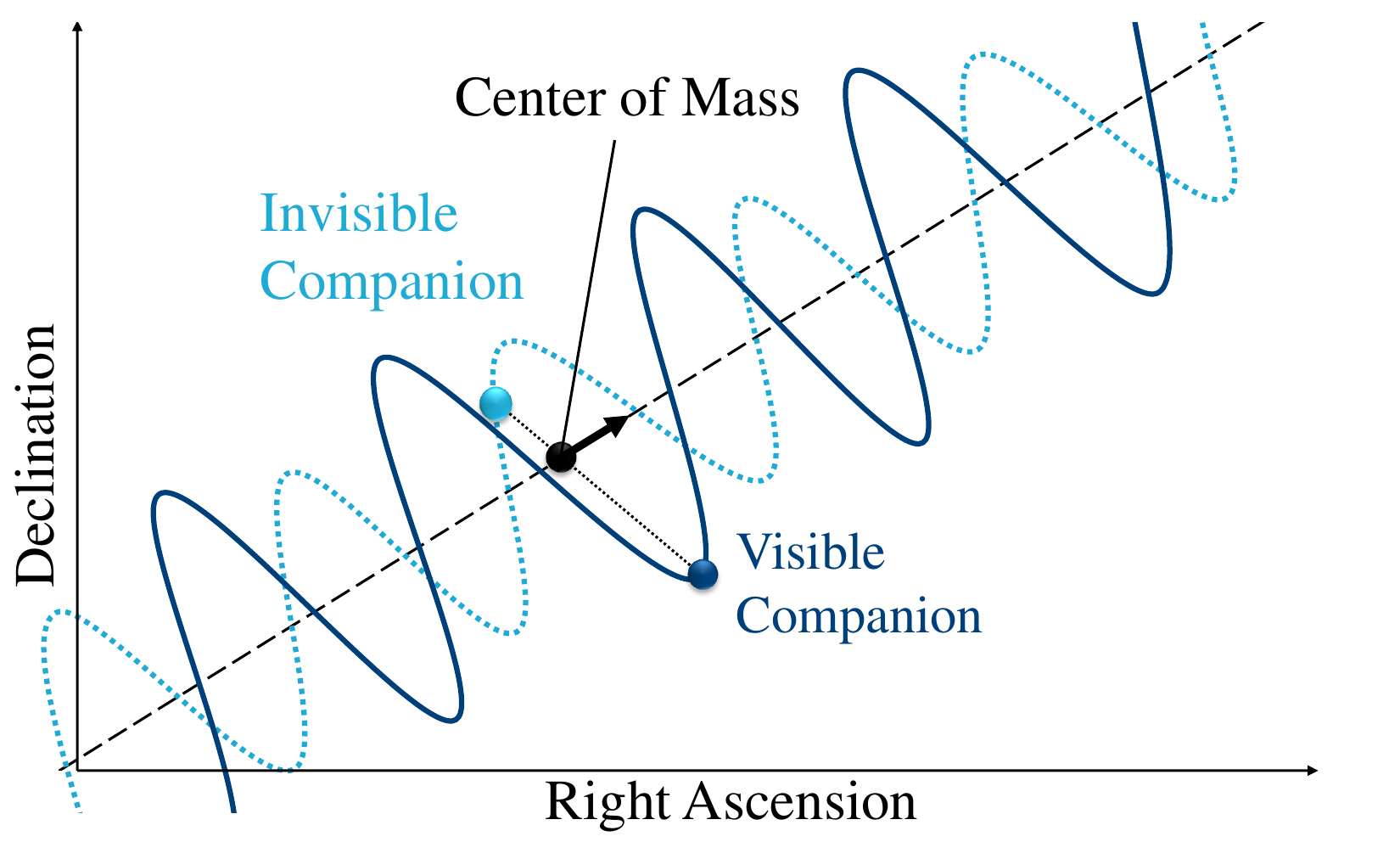}
\caption{(Left) RV curve of a circular and eccentric binary with $M_1=10$~\msun\ and $q=0.5$. (Right) Illustration of the effect of orbital motion on the apparent position of the visible companion star, as measured with astrometry. Due to the gravitational influence, the wobble in the orbit can be traced over time.}
\label{fig:astrometry}
\end{figure}

These equations can be rewritten to let the masses appear:
\begin{eqnarray}
    v_\mathrm{r,1}(t) - \gamma &=&  \left(\frac{2\pi}{P} \frac{\mathrm{G} M_2}{(1+1/q)^2} \right)^{1/3} \frac{\sin i}{\sqrt{1-e^2}}  \left[\cos (\vartheta_1(t) + \omega_1) + e \cos \omega_1\right], \label{eq:vr1} \\
    v_\mathrm{r,2}(t) - \gamma  &=&   - \left(\frac{2\pi}{P} \frac{\mathrm{G} M_1}{(1+q)^{2}} \right)^{1/3} \frac{\sin i}{\sqrt{1-e^2}} 
    \left[\cos (\vartheta_1(t) + \omega_1) + e \cos \omega_1\right], \label{eq:vr2}
\end{eqnarray}
where $q=M_2/M_1$ is the mass ratio and where we have used $\omega_2 = \omega_1 - \pi$.
These equations show that short orbital periods, large inclinations and higher masses will lead to higher amplitudes of the RV variations. Pole-on systems ($i=0$) will yield no RV variations so that low-inclination systems will not reveal themselves through spectroscopy. A relation of interest is further obtained by dividing the left- and right-hand terms of Eq.~\eqref{eq:vr2} by those of Eq.~\eqref{eq:vr1}. After some algebra, one obtains:
\begin{equation}
    v_\mathrm{r,2}(t)=-v_\mathrm{r,1}(t)/q +\gamma (1+1/q). \label{eq:v2ov1}
\end{equation}
Given a sufficient set of simultaneous RV measurements of both stars, the slope of Eq.~\eqref{eq:v2ov1} provides a direct estimate of the mass-ratio without the need for an orbital solution. It also nicely demonstrates that the instantaneous  (radial) velocities  of both stars are anti correlated (Fig.~\ref{fig:astrometry}, left) and that double-lined spectroscopic binary systems provide model-independent access to the mass-ratio.

We have so far written the radial velocity equations of both components. However, both components may not necessarily be detectable. This is often the case if one component is significantly fainter than its companion so that its spectral lines are lost in the noise of the spectrum. Systems where the signature of both components are seen in the combined spectrum are called {\bf double-lined spectroscopic binaries (SB2)}. Systems whose spectra only reveal one component are called {\bf single-lined spectroscopic binaries (SB1)}.
As a rule of thumb, main-sequence systems detected as SB2s have a brightness ratio $f_2/f_1>0.1$, or approximately $q>0.5$ given a typical mass-luminosity relation approximation of $M\propto L^3$. Modern analysis techniques, such as spectral disentangling, allow extracting the average spectra of both companions down to much less favorable flux ratio ($f_2/f_1\approx0.002$, and mass ratio nearing $q=0.1$). 

Detecting the signature of both companions is clearly advantageous. As shown above, this provides a direct constraint on the system mass ratio as $q=M_2/M_1=K_1/K_2$. It also allows constraining the minimum mass of each component:
\begin{eqnarray}    
    M_1 \sin^3 i &=& 1.0361\times 10^{-7} P K_2^3 (1+1/q)^2 (1-e^2)^{3/2},   \label{eq:m1sin3i}\\
    M_2 \sin^3 i&=& 1.0361\times 10^{-7} P K_1^3 P (1+q)^2  (1-e^2)^{3/2}.   \label{eq:m2sin3i}
\end{eqnarray}
where $M_{1,2}$ is in solar mass, $P$ in days, and $K_{1,2}$ in \kms.

For SB1 systems, only a minimum mass  on the secondary, unseen component can be obtained through the so-called mass function:

\begin{equation}
    f_{M_2}=\frac{(M_2 \sin i)^3}{(M_1+M_2)^2}=\frac{M_2 \sin^3 i}{1+1/q}=\frac{PK_1^3}{2 \pi G}\leq M_2. \label{eq:mass_funct}
\end{equation}
Given $f_{M_2} < M_2 \sin^3 i $, the constraints from Eq.~\eqref{eq:mass_funct} are not as informative as that from Eq.~\eqref{eq:m2sin3i} and, in most cases, the mass function largely underestimates $M_2$.

\subsection{Astrometric binaries} \label{s:astrometry}
Astrometry is concerned with the precise measurement of the angular positions of objects on the celestial sphere, as well as the variations of these positions over time. For example, astrometry allows measuring the proper motion of celestial objects, that is the angular speed of the trajectory of their photocenter on the plane of the sky. In the absence of any forces (for example, for single, isolated stars unaffected by the galactic potential), the proper motion of the star is expected to be constant and the object to move on a straight line over time. For binary systems formed by unequally bright objects, the situation is different. Even in the extreme case where only one star is visible, it is possible to identify it as a binary system. Due to the presence of the companion, the observed star will move around the center of mass of the system, which will cause a wobble in the trajectory of the center of light, as shown in the sketch of Fig. \ref{fig:astrometry} (right).

\begin{figure}[t]
    \centering
    \includegraphics[width=0.95\linewidth]{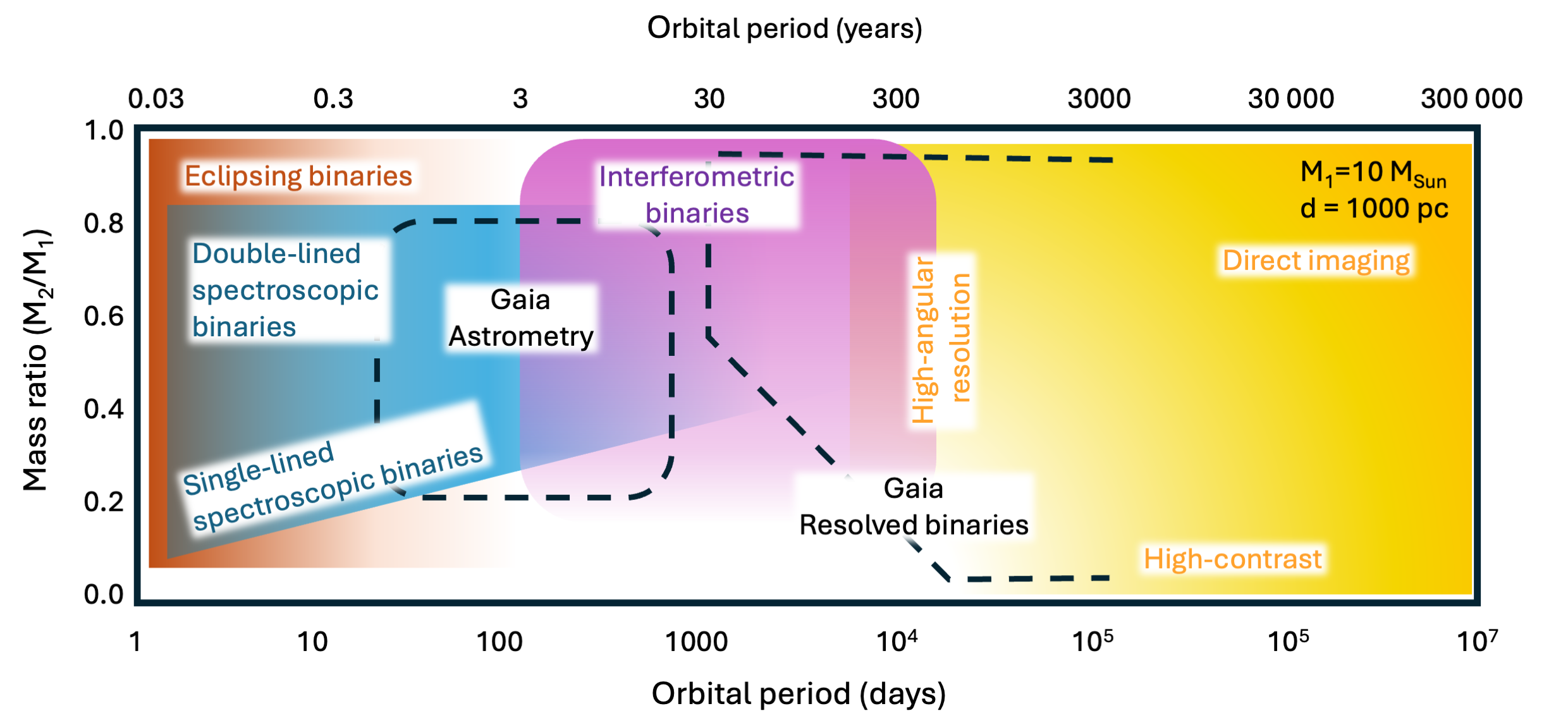}
    \caption{Approximate coverage of the period-mass ratio parameter space for a 10~\msun\ primary star located a 1000~pc. Areas of relevance of a subset of techniques have been indicated.}
    \label{f:param_space}
\end{figure}

Astrometry can also detect binaries even if it does not resolve them, that is, the angular separation between the components $\rho$ stays below the resolution capabilities of the telescope. In this situation, astrometric measurements will then trace the wobbling of the photocenter of the system.
%\begin{equation}
%    \rho<\theta_\mathrm{min}\approx1.22\frac{\lambda}{\phi},
%\end{equation}
%where $\phi$ is the diameter of the telescope aperture,  $\lambda$ the wavelength of observations, and $\theta_\mathrm{min}$ the diffraction limit of the optical system. 
While it is a clear signature of multiplicity, we note that the wobbling of the photocenter is not directly representative of the motion of the two stars but depends on the light-ratio. The semi-major axis of the apparent orbit or the photocenter ($a_0$) is therefore not related to the true masses by Eq.~\eqref{eq:K3} anymore, and an orbital fit requires treating $a_0$ as an independent free parameter (hence using seven orbital parameters). However, the trajectory of the photocenter still provides direct access to orbital quantities such as the orbital period and geometry of the orbit. With knowledge of the light ratio through, e.g., a mass-luminosity relation, one can further estimate the masses of the parameters in the systems. As for interferometric binaries, astrometric binaries are easier to detect if they are closer and have longer orbital periods.  The observations must often span a considerable amount of time to map the changes in the proper motion.  The ESA Gaia space mission is probably the most advanced astrometric mission so far \citep{gaia}. Upon completion, it will have scanned the position of $\sim$one billion objects over a decade, covering a large fraction of the binary discovery space (Fig.~\ref{f:param_space}; see also \citeauthor{ElBadry2024} \citeyear{ElBadry2024}).

\begin{table}[t]
    \centering
    \caption{Overview of parameters for which model-independent constraints can be obtained by various techniques, or a combination thereof. }
    \label{t:solutions}
    \begin{tabular}{c c c c c c}
    \hline
        &   \multicolumn{5}{c}{Techniques}\\
    \hline
        & Double-  &       & Double-lined  &          & Relative    \\
        & Eclipses & EB2    & Spectr. Bin  &   SB2+   & Astrometry  \\
Quantity& (EB2)    &  +SB2 & (SB2)         &  rAB    & (rAB)        \\
\hline
$P$     & $\checkmark$ & $\checkmark$ & $\checkmark$ & $\checkmark$ & $\checkmark$ \\
$e$     & $\checkmark$ & $\checkmark$ & $\checkmark$ & $\checkmark$ & $\checkmark$ \\
$T$     & $\checkmark$ & $\checkmark$ & $\checkmark$ & $\checkmark$ & $\checkmark$ \\
$\omega$& $\checkmark$ & $\checkmark$ & $\checkmark$ & $\checkmark$ & $\checkmark$ \\
$i$     & $\checkmark$ & $\checkmark$ & $\times$     & $\checkmark$ & $\checkmark$ \\
$\Omega$& $\times$     & $\checkmark$ & $\times$     & $\checkmark$ & $\checkmark$ \\
$q$     & $\times$     & $\checkmark$ & $\checkmark$ & $\checkmark$ & $\times$ \\
$d$     & $\times$     & $\checkmark$ & $\times$     & $\checkmark$ & $\times$ \\
Size    &  $R_{1,2}/a_{1,2}$     & $a_{1,2}$, $R_{1,2}$  & $a_{1,2}\sin i$ & $a_{1,2}$ & $a_\mathrm{tot}/d$\\
Mass  &  $\approx M_{1,2}/R_{1,2}$ & $M_{1,2}$   & $M_{1,2}\sin^3 i$&$M_{1,2}$&$M_\mathrm{tot}$\\
    \hline
    \end{tabular}
\end{table}
\subsection{Interferometric binaries} \label{s:interfero}

Interferometry is a technique that coherently recombines the light collected from different telescopes, allowing to achieve an angular resolution $\theta=1.22 \lambda/D$, defined not by the diameter $\phi$ of an individual telescope as is the case in single-dish observations, but by  the separation $D$ between the telescopes or antenna. Separations of several 100~m are used in optical-IR long baseline interferometry (OLBIN), while baselines of 10 to 1000s of km are used in sub-mm and radio-interferometry. Except for the early formation and evolved, dusty phases, stellar systems are mostly studied with OLBIN. OLBIN observables are visibilities, closure phases and differential phases  that can either be analyzed using forward modeling, or by applying image reconstruction if the data are sufficiently covering the spatial frequencies \citep{Buscher2015}. Modern instruments also provide spectral capabilities \citep[e.g][]{Eisenhauer2023}. We cannot provide here a full description on how to analyze interferometric observations, but it has to suffice here to say that, in the specific case of a binary system with two optical bright companions, interferometry provides a measurement of the angular separation $\rho$ between the companions and the position angle with respect to North (alternatively, $\delta N$, $\delta E$). Interferometry also constrains the flux ratio in the observed band. The instantaneous angular separation $\rho$ is a projection of the true instantaneous separation on the plane of the sky. It depends in a non-trivial way on the distance $d$ to the object, the orientation of the orbit ($i, \omega$, $\Omega$) and the moment ($\vartheta$) at which the pair is observed in its orbital motion:
\begin{equation}
\rho(t)=\frac{a}{d}\frac{(1-e^2)}{1+e\cos\vartheta(t)} \left[\cos^2\left( w+\vartheta(t)\right)+\sin^2\left( w+\vartheta(t)\right) \cos^2 i \right]^{1/2}. \label{eq:rho}
\end{equation}

The best-fit orbital solution is usually obtained numerically by solving the Thiele-Innes constants ($A, B, F, G$):
\begin{eqnarray}
    A = a [\cos \Omega \cos \omega - \sin \Omega \sin \omega \cos i], \hspace*{1cm}& F = a[-\cos \Omega \sin \omega-\sin\Omega\cos \omega \cos i],\\
    B = a[\sin \Omega \cos \omega + \cos \Omega \sin \omega \cos i], \hspace*{1cm}& 
    G =  a [- \sin \Omega \sin \omega + \cos \Omega \cos \omega \cos i], 
\end{eqnarray}
so that 
\begin{eqnarray}
\delta N& = &BX+GY,\\
\delta E& = &AX+FY.
\end{eqnarray}
Given enough data points, one can then fit the relative astrometric orbit. Beside the orbital parameters $P, e, \omega$ and $T$, a relative astrometric orbit gives specifically access to  the orbital inclination $i$ and total mass $M$ of the system with often excellent accuracy.

\subsection{High-angular resolution and high-contrast imaging}
Most of the techniques covered so far observed the binaries in integrated light. {\bf High-angular resolution imaging} techniques however allow us to detect and directly image both stars independently. This requires the angular separation $\rho$ to be larger than the angular resolution $\theta_\mathrm{min}=1.22 \lambda/\phi$ of the telescope. Observations from the ground are further affected by atmospheric turbulence that blurs the image. With some of the best observing site delivering seeing quality of 0.5" to 1" it is straightforward to estimate that ground-based optical observations with telescopes larger 10 or 20~cm  will be limited by atmospheric turbulence rather than by the theoretical diffraction limit of the telescope. 

To limit the impact of turbulence, modern instrumentation dedicated to high-angular resolution imaging are often equipped with {\bf adaptive optics}, a method that tracks the atmospheric distortion of the image in real-time and corrects for it through a deformable mirror inserted into the light path \citep{ao2012}. Alternative methods can be used in combination with, or without, adaptive optics. This includes {\bf speckle imaging}, where the exposure time is reduced to a fraction of a second, allowing to freeze the effect of turbulence at the cost of integrated flux levels per exposure. Post-processing of the images including re-centering and a selection of the best frames, allow astronomers to recover the theoretical diffraction limit of the telescope.

Besides the difficulty to resolve a pair of stars, it is also challenging to detect a faint companion near a bright star. The set of techniques dedicated to solving this challenge is broadly called {\bf high-contrast imaging} \citep{Oppenheimer2009}. The main challenge that high-contrast imaging needs to solve is to distinguish the presence of (faint) companions in the photon noise of the wings of the point-spread function of a much brighter central star. Nowadays, high-contrast imaging also strives for high-angular resolution performances as the latter allows for the light of the central object to be more spatially concentrated, hence the light from the bright companion {\it leaks} less onto the position of the fainter one \citep[e.g.,][]{xao2018}. 

In addition to adaptive optics, {\bf coronagraphy} is one of the most popular techniques used to attenuate the light of the central star. It consists of inserting a physical mask in the light path that blocks the light of the bright companion. Combined with suitable image post-processing, flux contrasts better than 1:100\, 000 (magnitude difference $>10$) are routinely obtained at separations of a few times the diffraction limit.  

{\bf Aperture masking} can be seen as a hybrid technique which transforms a single dish in a set of multi-dish, small-aperture telescopes  by inserting a physical mask that blocks the pupil plane, except for a set of holes in the mask \citep{Lacour2011,Ireland2013}. Given the light coherently passes through these holes, they produced a speckle interferograph on the detector that can then be analyzed using a similar approach as for interferometry, such as forward modeling of the speckle patterns in the spatial or Fourier space, or, image reconstruction. 

In all these advanced imaging techniques, as in the case of classical, seeing-limited imaging, the information sought is very similar to those introduced above.  Often, astronomers measure the relative angular position and brightness contrast of the companions to the central sources (e.g., Sect.~\ref{s:interfero}). In some cases, absolute position (as in e.g., Sect.~\ref{s:astrometry}) and/or fluxes are considered depending on the science goal and the properties of the system. While probing larger and larger separations for a given set of masses and separations, the orbital periods increase consequently so that it soon becomes challenging to detect any signature of the orbital motion and statistical methods then need to be used.

\subsection{Parameter space and multi-technique observations}

Different methods can be used to  unveil the binary nature of an object, and to characterize its orbital and physical parameters. These techniques  will be more or less successful (or not at all) depending on the observational properties of the binary systems and the characteristics of the observational campaign, including the precision and frequency of the measurements.  When trying to assess the number of binaries in a given population, it is generally useful to combine different techniques in order to better map the parameter space that binaries cover and Fig.~\ref{f:param_space} provides a simplified view of the sensitivity of various techniques. In certain regions of the parameter space, binaries can be studied by different techniques. As summarized in Table~\ref{t:solutions}, combining different techniques is a must if one is to derive model-independent quantities such as distance, masses and absolute separations.

\section{Conclusions}

Since the first detection of the orbital motion of a binary system just over 200 years ago, these objects have formed a privileged laboratory to progress  stellar physics. For the first time, astronomers were able to measure and weight stars, which was a key step in exploring fundamental stellar physics through relations such as the mass-luminosity and mass-radius relations \citep{Torres2010}. 
We have shown here how to use some of the observational properties of binary stars to retrieve these physical properties. Other detection methods exist, based on other physical phenomena (e.g., polarimetry, Doppler beaming, colliding winds) and/or other wavelength domains (e.g., radio, X-rays, gravitational waves).  We also barely scratched the surface of the statistical methods which, in today's big data science, represent a significant shortcoming of this article.

The study of binaries and higher order multiples also goes far beyond these fundamental quantities. The presence of two stars in close proximity leads to novel physics, involving complex interplays between the stars and new evolution pathways that dramatically alter the end-of-life products of the stars involved.  We have indeed avoided any specifics related to stellar (binary) evolution and the diversity of intriguing objects it produces, including binaries with stripped stars, critical rotators, inert compact objects, X-ray binaries, pulsar binaries and gravitational wave sources. We encourage the interested readers to pursue their journey by browsing for these topics in the present encyclopedia. We also provide below a select list of recent review articles and textbook on the topics.

\bibliographystyle{Harvard}
\bibliography{main}

\end{document}